\documentclass[]{article}

\usepackage{mysprocl}
\usepackage{cite}
\usepackage{epsfig}
\def\beq{\begin{equation}}
\def\eeq{\end{equation}}
\def\beqar{\begin{eqnarray}}
\def\eeqar{\end{eqnarray}}
\def\barr#1{\begin{array}{#1}}
\def\earr{\end{array}}
\def\bfi{\begin{figure}}
\def\efi{\end{figure}}
\def\btab{\begin{table}}
\def\etab{\end{table}}
\def\bce{\begin{center}}
\def\ece{\end{center}}

\def\text{\textstyle}

\def\al{\alpha}

\def\De{\Delta}

\def\refeq#1{\mbox{eq.~(\ref{#1})}}
\def\refeqs#1{\mbox{eqs.~(\ref{#1})}}
\def\reffi#1{\mbox{Fig.~\ref{#1}}}

\def\citere#1{\mbox{Ref.~\cite{#1}}}
\def\citeres#1{\mbox{Refs.~\cite{#1}}}

\def\mathswitchr#1{\relax\ifmmode{\mathrm{#1}}\else$\mathrm{#1}$\fi}

\newcommand{\PZ}{\mathswitchr Z}
\newcommand{\PA}{\mathswitchr A}

\newcommand{\PH}{\mathswitchr H}

\newcommand{\Ph}{\mathswitchr h}

\newcommand{\Pt}{\mathswitchr t}

\def\mathswitch#1{\relax\ifmmode#1\else$#1$\fi}

\newcommand{\MZ}{\mathswitch {M_\PZ}}

\newcommand{\Mt}{\mathswitch {m_\Pt}}

\newcommand{\mh}{\mathswitch {m_\Ph}}
\newcommand{\mH}{\mathswitch {m_\PH}}
\newcommand{\MA}{\mathswitch {M_\PA}}

\newcommand{\GF}{\mathswitch {G_\mu}}
\newcommand{\gf}{\GF}

\def\tb{\tan\beta}

\newcommand{\Mstr}{M_{\tilde{t}_R}}
\newcommand{\Mstl}{M_{\tilde{t}_L}}

\newcommand{\Xt}{X_{\Pt}}

\newcommand{\lmtmsms}{\KL\frac{\mtms^2}{\ms^2}\KR}

\newcommand{\mt}{\Mt}
\newcommand{\mtms}{\overline{m}_{\Pt}}
\newcommand{\mtbar}{\mtms}

\newcommand{\Msbar}{\overline{M}_{\mathrm{S}}}
\newcommand{\Xtbar}{\overline{X}_{\Pt}}

\newcommand{\mgl}{m_{\tilde{\mathrm{g}}}}

\newcommand{\tsf}{\theta\kern-.20em_{\tilde{f}}}
\newcommand{\tsfp}{\theta\kern-.20em_{\tilde{f}\prime}}
\newcommand{\tsq}{\theta\kern-.15em_{\tilde{q}}}

\newcommand{\msusy}{M_{\mathrm{SUSY}}}
\newcommand{\ms}{M_{\mathrm{S}}}

\newcommand{\lsim}
{\;\raisebox{-.3em}{$\stackrel{\displaystyle <}{\sim}$}\;}

\newcommand{\fh}{{\em FeynHiggs}}

\newcommand{\msbar}{$\overline{\rm{MS}}$}

\newcommand{\oas}{{\cal O}(\alpha_s)}
\newcommand{\oaas}{{\cal O}(\alpha\alpha_s)}
\newcommand{\cp}{{\cal CP}}
\newcommand{\wz}{\sqrt{2}}

\newcommand{\twol}{two-loop}
\newcommand{\onel}{one-loop}

\newcommand{\KL}{\left(}
\newcommand{\KR}{\right)}

\newcommand{\VL}{\left( \begin{array}{c}}
\newcommand{\VR}{\end{array} \right)}
\newcommand{\ML}{\left( \begin{array}{cc}}
\newcommand{\MLd}{\left( \begin{array}{ccc}}
\newcommand{\MLv}{\left( \begin{array}{cccc}}
\newcommand{\MR}{\end{array} \right)}

\newcommand{\gev}{\,\, \mathrm{GeV}}

\newcommand{\BC}{\begin{center}}
\newcommand{\EC}{\end{center}}
\newcommand{\BE}{\begin{equation}}
\newcommand{\EE}{\end{equation}}
\newcommand{\BEA}{\begin{eqnarray}}
\newcommand{\BEAnn}{\begin{eqnarray*}}
\newcommand{\EEA}{\end{eqnarray}}
\newcommand{\EEAnn}{\end{eqnarray*}}

\newcommand{\id}{{\rm 1\kern-.12em
\rule{0.3pt}{1.5ex}\raisebox{0.0ex}{\rule{0.1em}{0.3pt}}}}

\def\als{\alpha_s}

\def\hSi{\hat{\Sigma}}

\def\draftdate{\relax}
\def\mda{\relax}
\def\mua{\relax}
\def\mla{\relax}
\def\draft{
\def\thtystars{******************************}
\def\sixtystars{\thtystars\thtystars}
\typeout{}
\typeout{\sixtystars**}
\typeout{* Draft mode!
         For final version remove \protect\draft\space in source file
*}
\typeout{\sixtystars**}
\typeout{}
\def\draftdate{\today}
\def\mua{\marginpar[\boldmath\hfil$\uparrow$]%
                   {\boldmath$\uparrow$\hfil}%
                    \typeout{marginpar: $\uparrow$}\ignorespaces}
\def\mda{\marginpar[\boldmath\hfil$\downarrow$]%
                   {\boldmath$\downarrow$\hfil}%
                    \typeout{marginpar: $\downarrow$}\ignorespaces}
\def\mla{\marginpar[\boldmath\hfil$\rightarrow$]%
                   {\boldmath$\leftarrow $\hfil}%
                    \typeout{marginpar:
$\leftrightarrow$}\ignorespaces}
\def\Mua{\marginpar[\boldmath\hfil$\Uparrow$]%
                   {\boldmath$\Uparrow$\hfil}%
                    \typeout{marginpar: $\Uparrow$}\ignorespaces}
\def\Mda{\marginpar[\boldmath\hfil$\Downarrow$]%
                   {\boldmath$\Downarrow$\hfil}%
                    \typeout{marginpar: $\Downarrow$}\ignorespaces}
\def\Mla{\marginpar[\boldmath\hfil$\Rightarrow$]%
                   {\boldmath$\Leftarrow $\hfil}%
                    \typeout{marginpar:
$\Leftrightarrow$}\ignorespaces}
\overfullrule 5pt
\oddsidemargin -15mm
\marginparwidth 29mm
}

\begin{document}

\thispagestyle{empty}

\null
\hfill KA-TP-16-1999\\
\null
\hfill DESY 99-148\\
\null
\hfill hep-ph/9910283\\
\vskip .8cm
\begin{center}
{\Large \bf Precise Calculations for the Neutral\\[.5em]
Higgs-Boson Masses in the MSSM
\footnote{Presented by G.~Weiglein at the International Workshop on Linear
Colliders, Sitges,\\
April~28 -- May 5, 1999.}
}
\vskip 2.5em
{\large
{\sc S.\ Heinemeyer$^1$, W.\ Hollik$^{2}$ and G.\ Weiglein$^3$}\\[1em]
{\normalsize \it $^1$ DESY Theorie, Notkestr. 85, D--22603 Hamburg,
Germany}\\[.3em]
{\normalsize \it $^2$ Institut f\"ur Theoretische Physik, Universit\"at
Karlsruhe,\\
D--76128 Karlsruhe, Germany}\\[.3em]
{\normalsize \it $^3$ CERN, TH Division, CH--1211
Geneva 23, Switzerland}
}
\vskip 2em
\end{center} \par
\vskip 1.2cm
\vfil
\bce
{\bf Abstract} \par
\ece
We review the comparison of the results for the neutral 
$\cp$-even Higgs-boson masses recently obtained within the
Feynman-diagrammatic (FD) approach with the previous results based on
the renormalization group (RG) approach. We show that the results differ
by new genuine \twol\ contributions present in the FD calculation. The
numerical effect of these terms on the result for $\mh$ is briefly
discussed.
\par
\vskip 1cm
\null
\setcounter{page}{0}
\clearpage

\title{PRECISE CALCULATIONS FOR THE NEUTRAL\\
       HIGGS-BOSON MASSES IN THE MSSM
}

\author{S.~HEINEMEYER}
\address{DESY Theorie, Notkestr. 85, D--22603 Hamburg, Germany}
\author{W.~HOLLIK, G.~WEIGLEIN
\footnote{Address after 30.09.99: CERN, TH Division, CH--1211 Geneva 23,
Switzerland}
}
\address{Institut f\"ur Theoretische Physik, 
         Universit\"at Karlsruhe, \\ 
         D--76128 Karlsruhe, Germany}
%
%
\maketitle
\abstracts{
We review the comparison of the results for the neutral 
$\cp$-even Higgs-boson masses recently obtained within the
Feynman-diagrammatic (FD) approach with the previous results based on
the renormalization group (RG) approach. We show that the results differ
by new genuine \twol\ contributions present in the FD calculation. The
numerical effect of these terms on the result for $\mh$ is briefly
discussed.
}


\section{Theoretical basis}

Within the Minimal Supersymmetric Standard Model (MSSM) the masses of
the $\cp$-even neutral Higgs bosons are 
calculable in terms of the other MSSM parameters. The mass of the
lightest Higgs boson, $\mh$, has been of particular interest as it is
bounded from above at the tree level to be smaller than the Z-boson
mass. This bound, however, receives large radiative corrections. The \onel\
results~\cite{mhiggs1l,mhiggsf1l,mhiggsf1lb} have been
supplemented in the last years with the leading \twol\ corrections, 
performed in the renormalization group (RG)
approach~\cite{mhiggsRG}, in the effective
potential approach~\cite{mhiggsEP} and most recently in
the Feynman-diagrammatic (FD) approach~\cite{mhiggsFD}.
These calculations predict an
upper bound on $\mh$ of about $\mh \lsim 135 \gev$.

The dominant radiative corrections to $\mh$ arise from the top and
scalar top sector of the MSSM, with the input parameters $\mt$, $\msusy$ and 
$\Xt$. Here we assume the soft SUSY breaking parameters in the diagonal 
entries of the scalar top mixing matrix to be equal for simplicity,
$\msusy = \Mstl = \Mstr$. The off-diagonal entry of the mixing
matrix in our conventions (see \citere{mhiggsFD}) reads 
$\mt \Xt = \mt (A_t - \mu \cot\beta)$. We furthermore use the short-hand 
notation $\ms^2 := \msusy^2 + \mt^2$.

Within the RG approach, $\mh$ is calculated from the effective
renormalized Higgs quartic coupling at the scale $Q = \mt$. The RG
improved leading logarithmic approximation is obtained by applying the 
one-loop RG running of this coupling from the high scale $Q = \ms$ 
to the scale $Q = \mt$ and including the one-loop threshold effects from
the decoupling of the supersymmetric particles at $\ms$~\cite{mhiggsRG}. 
This approach
relies on using the \msbar\ renormalization scheme. The parameters in
terms of which the RG result for $\mh$ is expressed are thus 
\msbar\ parameters.

In the FD approach, the masses of the $\cp$-even Higgs bosons are determined
by the poles of the corresponding propagators. The corrections to the
masses $\mh$ and $\mH$ are thus obtained by evaluating loop corrections
to the $h$, $H$ and $hH$-mixing propagators. The poles of the
corresponding propagator matrix are given by the solutions of 
\BE
\label{eq:mhpole}
\left[q^2 - m_{\Ph, {\rm tree}}^2 + \hSi_{hh}(q^2) \right]
\left[q^2 - m_{\PH, {\rm tree}}^2 + \hSi_{HH}(q^2) \right] -
\left[\hSi_{hH}(q^2)\right]^2 = 0 ,
\EE
where $\hSi_{hh}(q^2)$, $\hSi_{HH}(q^2)$,
$\hSi_{hH}(q^2)$ denote the renormalized Higgs-boson
self-energies. In \citere{mhiggsFD} the dominant two-loop
contributions to the masses of the $\cp$-even Higgs bosons of $\oaas$ 
have been evaluated. These corrections, obtained in the on-shell
scheme, have been combined with the complete one-loop
on-shell result~\cite{mhiggsf1lb} and the sub-dominant two-loop
corrections of ${\cal O}(\gf^2 \mt^6)$~\cite{mhiggsRG}. The
corresponding results have been implemented into the Fortran code
\fh~\cite{feynhiggs}.


\section{Leading two-loop contributions in the FD approach}

In \citere{mhiggslle} the leading contributions have been extracted via
a Taylor expansion from the rather complicated diagrammatic two-loop
result obtained in \citere{mhiggsFD} and a compact expression for
the dominant contributions has been derived. Restricting to the
leading terms in $\mt/\ms$, $\MZ^2/\mt^2$ and $\MZ^2/\MA^2$, the
expression up to $\oaas$ reduces to the simple form
\BE
m_{\Ph, \mathrm{FD}}^2 = \mh^{2,{\rm tree}} +
\De m_{\Ph, \mathrm{FD}}^{2,\al} +
\De m_{\Ph, \mathrm{FD}}^{2,\al\als},
\label{eq:resFD}
\EE
where the one-loop correction is given by
\BE
\De m_{\Ph, \mathrm{FD}}^{2,\al} =   \frac{3}{2} \frac{\gf
\sqrt{2}}{\pi^2} \mtbar^4 \left\{
  - \ln\left(\frac{\mtbar^2}{\ms^2} \right)
  + \frac{\Xt^2}{\ms^2}
  \left(1 - \frac{1}{12} \frac{\Xt^2}{\ms^2} \right)
  \right\} .
\label{eq:oneloop}
\EE
The two-loop contribution reads
\BEA 
\label{mh2twolooptop}
\De m_{\Ph, \mathrm{FD}}^{2,\al\als} &=& \De m_{\Ph,\rm log}^{2,\al\als} 
                     + \De m_{\Ph,\rm non-log}^{2,\al\als} , \\
\De m_{\Ph,\rm log}^{2,\al\als} &=&
    - \frac{\gf\wz}{\pi^2} \frac{\als}{\pi}\; \mtms^4
      \left[ 3 \ln^2\lmtmsms + \ln\lmtmsms 
     \KL 2 - 3 \frac{\Xt^2}{\ms^2} \KR \right] , 
\label{eq:mhlog} \\
\De m_{\Ph,\rm non-log}^{2,\al\als} &=& 
    - \frac{\gf\wz}{\pi^2} \frac{\als}{\pi}\; \mtms^4
      \left[ 4 -6 \frac{\Xt}{\ms} 
     - 8 \frac{\Xt^2}{\ms^2} 
     +\frac{17}{12} \frac{\Xt^4}{\ms^4} \right] ,
\label{eq:mhnolog}
\EEA
in which the leading logarithmic and the
non-logarithmic terms have been given separately. The parameter
$\mtms$ in \refeqs{eq:oneloop}--(\ref{eq:mhnolog}) denotes the running
top-quark mass at the scale $\mt$, which is related to the pole mass
$\mt$ in $\oas$ via
\BE
\mtms \equiv \mtms(\mt) =
\frac{\mt}{1 + \frac{4}{3\,\pi} \als(\mt)} ,
\label{mtrun}
\EE
while $\ms$ and $\Xt$ are on-shell parameters. 

\begin{figure}[ht!]
\vspace{1em}
\begin{center}
\mbox{
\psfig{figure=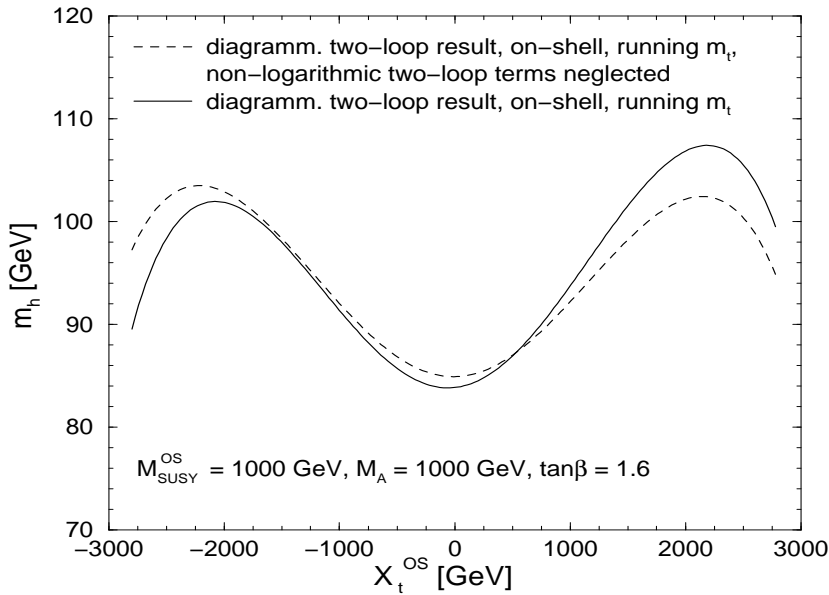,width=11cm,height=8.0cm}}
\end{center}
\caption[]{The dominant one-loop and two-loop contributions to 
$\mh$ evaluated in the FD approach are shown as a function of 
(the on-shell parameter) $\Xt$ for $\tb = 1.6$.
The full curve shows the result including the new genuine \twol\
contributions, \refeq{eq:mhnolog},
while the dashed curve shows the result where these
non-logarithmic two-loop corrections have been neglected.
}
\label{fig:mhnolog}
\end{figure}

The one-loop correction, \refeq{eq:oneloop}, as well as the 
dominant logarithmic two-loop contributions, \refeq{eq:mhlog}, are 
seen to be symmetric with respect to the sign of $\Xt$. 
The non-logarithmic two-loop
contributions, on the other hand, give rise to an asymmetry in the $\Xt$
dependence through the term in \refeq{eq:mhnolog} which is linear in $\Xt/\ms$.
The numerical effect of the non-logarithmic two-loop terms is
investigated in \reffi{fig:mhnolog}. The result for the dominant
contributions to $\mh$ of \refeqs{eq:resFD}--(\ref{eq:mhnolog})
is compared to the result where the
non-logarithmic contributions of \refeq{eq:mhnolog} are omitted. The
numerical effect of the non-logarithmic genuine two-loop contributions
is seen to be sizable. Besides a considerable asymmetry in $\Xt$ the
non-logarithmic two-loop terms in particular lead to an increase in the
maximal value of $\mh$ of about 5~GeV.

\section{Comparison between the FD and the RG approach}

The results for the dominant contributions derived by FD methods can be
compared with the explicit expressions obtained within the RG approach
which have been given in \citeres{mhiggsRG}. At the two-loop level,
the RG methods applied in \citeres{mhiggsRG} lead to the following result
in terms of the \msbar\ parameters $\mtbar$, $\Msbar$, $\Xtbar$
\BE 
\label{eq:mh2twoloopRG}
\De m_{\Ph, \mathrm{RG}}^{2,\al\als} =
    - \frac{\gf\wz}{\pi^2} \frac{\als}{\pi}\; \mtms^4
      \left \{3 \ln^2 \left(\frac{\mtbar^2}{\Msbar^2} \right) +
    \left[2 - 6 \frac{\Xtbar^2}{\Msbar^2}
          \left(1 - \frac{1}{12} \frac{\Xtbar^2}{\Msbar^2} \right) \right]
    \ln\left(\frac{\mtbar^2}{\Msbar^2} \right) \right \},
\EE
which solely consists of leading logarithmic contributions. The one-loop result
for the dominant contributions in the RG approach has the same form as
\refeq{eq:oneloop}, where the parameters $\ms$ and $\Xt$ have to be
replaced by $\Msbar$ and $\Xtbar$, respectively.

The one-loop RG-improved effective potential expression 
\refeq{eq:mh2twoloopRG} does not contain non-logarithmic contributions. 
In the viewpoint of the RG approach these genuine two-loop
contributions are interpreted as two-loop finite threshold corrections 
to the quartic Higgs couplings. 

For a direct comparison of the FD result given in the last section with
the RG result of \refeq{eq:mh2twoloopRG},
one has to take into account that $\ms$ and 
$\Xt$ in the FD result are on-shell parameters, 
while the corresponding parameters in the RG result, $\Msbar$ and $\Xtbar$, 
are \msbar\ quantities. The relations between these parameters are given in 
leading order by~\cite{bse} 
\BE 
\Msbar^2 = \ms^2
 - \frac{8}{3} \frac{\alpha_s}{\pi} \ms^2 , \qquad
\Xtbar = \Xt + \frac{\alpha_s}{3 \pi} \ms
   \left(8 + 4 \frac{\Xt}{\ms} - 
         3 \frac{\Xt}{\ms} \ln\left(\frac{\mt^2}{\ms^2}\right) \right) .
\label{eq:xtmsms} 
\EE 
Applying these relations for rewriting the FD result given in 
\refeqs{eq:resFD}--(\ref{eq:mhnolog}) in terms of the \msbar\ parameters
$\mtbar$, $\Msbar$, $\Xtbar$ one finds that the leading logarithmic
contributions in the two approaches in fact coincide~\cite{bse}, as it
should be as a matter of consistency. The FD result, however, contains 
further non-logarithmic genuine two-loop contributions which are not
present in the RG result. The effect of these extra terms within the 
\msbar\ parameterization considered here is qualitatively the same as
discussed in the preceding section. They lead to an asymmetry in 
the dependence of $\mh$ on $\Xt$ and to an increase in the maximal value
of $\mh$ compared to the RG result.


The analysis above has been performed for the dominant contributions
only. Further deviations between the RG and the FD result arise from 
non-leading one-loop and two-loop contributions, in which the results 
differ, and from varying the gluino mass, $\mgl$, in the FD result, which
does not appear as a parameter in the RG result. Changing the value of
$\mgl$ in the interval $0 \leq \mgl \leq 1$~TeV shifts the FD result 
relative to the RG result within $\pm 2$~GeV~\cite{mhiggsFD}.





\begin{thebibliography}{00}  



\bibitem{mhiggs1l} H.~Haber and R.~Hempfling,
                   {\em Phys. Rev. Lett.} {\bf 66} (1991) 1815;\\
                   J.~Ellis, G.~Ridolfi and F.~Zwirner,
                   {\em Phys. Lett.} {\bf B 257} (1991) 83; 
                   {\em Phys. Lett.} {\bf B 262} (1991) 477.

\bibitem{mhiggsf1l} P.~Chankowski, S.~Pokorski and J.~Rosiek,
                    {\em Nucl. Phys.} {\bf B 423} (1994) 437;
                    J.~Bagger, K.~Matchev, D.~Pierce, R.-J.~Zhang,
                    {\em Nucl. Phys.} {\bf B 491} (1997) 3.

\bibitem{mhiggsf1lb} A.~Dabelstein, 
                    {\em Nucl. Phys.} {\bf B 456} (1995) 25;
                    {\em Z. Phys.} {\bf C 67} (1995) 495.

\bibitem{mhiggsRG}  J.~Casas, J.~Espinosa, M.~Quir\'os and A.~Riotto,
                    {\em Nucl. Phys.} {\bf B 436} (1995) 3,
                    E: {\em ibid.} {\bf B 439} (1995) 466;\\
                    M.~Carena,~J.~Espinosa,~M.~Quir\'os,~C.~Wagner,
                    {\em Phys.~Lett.}~{\bf~B~355}~(1995)~209;\\
                    M.~Carena, M.~Quir\'os and C.~Wagner,
                    {\em Nucl. Phys.} {\bf B 461} (1996) 407;\\
                    H.~Haber, R.~Hempfling and A.~Hoang,
                    {\em Z. Phys.} {\bf C 75} (1997) 539.

\bibitem{mhiggsEP} R.~Hempfling and A.~Hoang,
                   {\em Phys. Lett.} {\bf B 331} (1994) 99;\\
                   R.-J.~Zhang, 
                   {\em Phys. Lett.} {\bf B 447} (1999) 89.

\bibitem{mhiggsFD} S.~Heinemeyer, W.~Hollik and G.~Weiglein,
                   {\em Phys. Rev.} {\bf D 58} (1998) 091701;
                   {\em Phys. Lett.} {\bf B 440} (1998) 296;
                   {\em Eur. Phys. Jour.} {\bf C 9} (1999) 343.


\bibitem{feynhiggs} S.~Heinemeyer, W.~Hollik and G.~Weiglein,
                    to appear in {\em Comp. Phys. Comm.}, 
                    hep-ph/9812320.

\bibitem{mhiggslle} S.~Heinemeyer, W.~Hollik and G.~Weiglein,
                    {\em Phys. Lett.} {\bf B 455} (1999) 179.

\bibitem{bse} M.~Carena, H.~Haber, S.~Heinemeyer, W.~Hollik, C.~Wagner
              and G.~Weiglein,
              {\em in preparation}.

\end{thebibliography}
\end{document}